\documentclass[9pt,twocolumn,twoside]{osajnl}

\journal{ol} 

\setboolean{shortarticle}{true} 

\title{Design of optical systems with toroidal curved detectors}

\author[1,2,*]{Eduard Muslimov}
\author[1]{Emmanuel Hugot}
\author[1]{Marc Ferrari}
\author[1]{Thibault Behaghel}
\author[1]{Gerard R. Lemaitre}
\author[1]{Melanie Roulet}
\author[1]{Simona Lombardo}

\affil[1]{Aix Marseille Univ, CNRS, CNES, LAM, Marseille, France}
\affil[2]{Kazan National Research Technical University named after A.N. Tupolev –KAI, 10 K. Marx, Kazan
420111, Russia}

\affil[*]{Corresponding author: eduard.muslimov@lam.fr}

\ociscodes{(040.1520) CCD, charge-coupled device, (220.1000) Aberration compensation, (220.1080) Active or adaptive optics, (110.6770) Telescopes.}

\begin{abstract}
We consider using  toroidal curved detectors to improve the performance of imaging optical systems. We demonstrate that some optical systems have an anamorphic field curvature. We consider an unobscured re-imaging three-mirror anastigmat as an example (\textit{f’}=960 mm, \textit{F/5.3}, $\omega_x$x$\omega_y$=4$^{\circ}$x4$^{\circ}$). By assuming that the image is focused on a toroidal detector surface and perform re-optimization, it becomes possible to obtain a notable gain in the image quality --- up to 40 \% in terms of the spot RMS radius.  Through analytic computations and finite-element analysis, we demonstrate that this toroidal shape can be obtained by bending of a thinned detector in a relatively simple setup.   
\end{abstract}

\setboolean{displaycopyright}{false}

\begin{document}

\maketitle

\section{Introduction}

Manufacturing of curved detectors is one of emerging technologies, which will define the performance of next generation optical instruments. \textbf{Since the topic was introduced for the first time  \citep{Nikzad01}, \citep{Nikzad10} and the early practical results with such detectors in imaging systems were shown \citep{Ko08}, their advantages have been explored in a number of publications (see for example \citep{Rim08}, \citep{Dumas12}). Recent use of curved detectors in large instruments like Space Surveillance Telescope (SST) \citep{Gregory15} has demonstrated the impact of these advantages on groundbreaking application.} With a curved detector there is no need to compensate for the Petzval curvature in focal planes. This implies that a certain number of degrees of freedom can be released to correct other aberrations. thus, a better image quality can be achieved. On the other hand, the optical design can be considerably simplified, with a decrease of the number of surfaces, their shape complexity, volume and weight of the system. On top of these advantages, the use of curved focal planes allows to gain in the image relative illumination.
A number of functional detectors were fabricated with use of different techniques like etching of silicon micro-springs \citep{Dinyari08}, bending of a thinned chip with an electromagnetic force \citep{Hatakeyama16}, mechanical force \citep{Tekaya14}, air pressure \citep{Guenter17} or thermal deformation \citep{Itonaga11}. 
\textbf{However, all of these developments considered only generation of a spherically shaped focal plane array.} Meanwhile, there is a large number of optical system for which the actual shape of the image surface is more complex. Moreover, in some cases this shape is anamorphic, i.e. it has no rotational symmetry.
Recently the authors explored the possibility of using toroidal detectors in a high-performance telescope \citep{Muslimov17}. This work represents a more systematic study of the topic.    
The main goal of the present paper is to demonstrate that by accounting for this anamorphic shape of the image surface, it is possible to achieve a notable gain in terms of the image quality. At the same time we will show that practical implementation of such a detector shape requires only a slight modification of the existing bending technique and doesn't introduce any significant complexity to the manufacturing process.
The paper is organized as follows: in section 2 we briefly describe the design strategy and algorithm; in section 3 we present a sample three-mirror anastigmat (TMA) design and show the image quality gain achievable with a toroidal focal plane; in section 4 the feasibility of this curved detector is demonstrated with finite element analysis; section 5 contains concluding remarks. 

\section{Design concept and algorithm}

The main idea of this study consists in using a toroidal detector shape to fit an anamorphic image surface. In order to show the potential advantages of this concept and its practical feasibility we perform the following steps:
\begin{enumerate}
  \item We consider an optical system containing \textbf{a deviation} from rotational symmetry and calculate the field curvature separately for the X and Y directions.
  \item The image surface is turned into a toroid. The values defined at the previous step  are assigned to the curvature radii; they are set as variables.
  \item The optical system is re-optimized with standard tools. The boundary conditions require maintenance of the main optical parameters and the principle geometry.
  \item For the found radii of curvature the corresponding shape of deformable substrate is computed. The X and Y sections are considered separately.
  \item A solid model of the substrate is created and tested with a finite-element analysis (FEA) technique. 
  \item The analysis is repeated for an assembly  including the substrate and the detector chip glued on the top. The mechanical load is adjusted iteratively to fit the required sag.
\end{enumerate}

\section{Example optical design}

Here we show the proposed design approach on an example of a re-imaging unobscured three mirror anastigmat (TMA) telescope \citep{Cook81},\citep{Gross05}. Such optical schemes provide relatively high image quality for an extended field of view and are widely used for different application. This particular design is notable for having a real exit pupil which is often required in infrared systems. 
Since the design is based on off-axis aspheres, the Petzval curvature should be different along the X and Y axes. Thus this case should fit our concept and below we demonstrate it in details.    

\subsection{Telescope design description}

\begin{figure}[h]
\centering
\fbox{\includegraphics[width=0.8\linewidth]{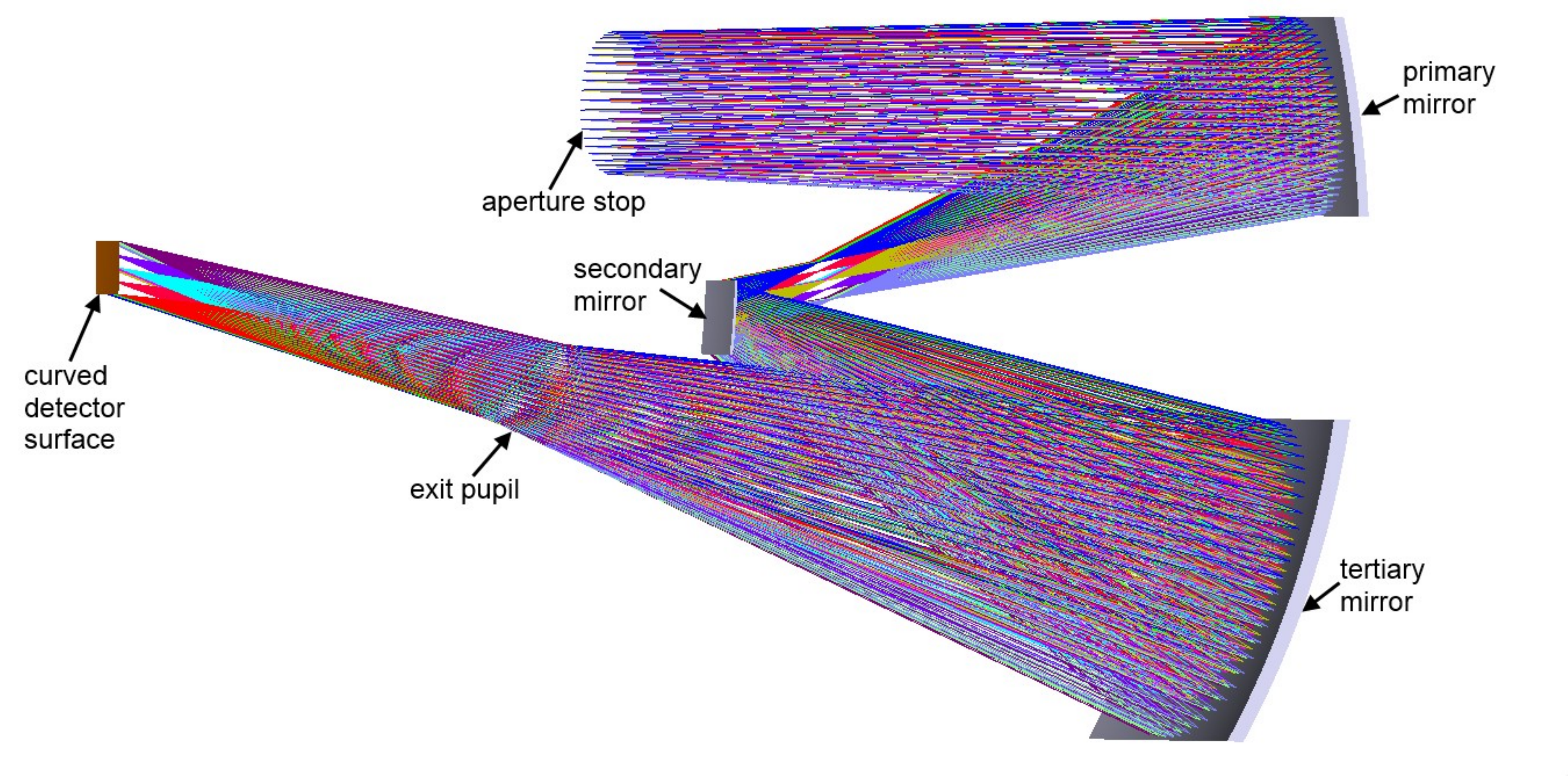}}
\caption{Optical scheme of a re-imaging unobscured TMA telescope ( \textit{f'}=960 mm, \textit{F/5.3}, 4$^{\circ}$x4$^{\circ}$). \textbf{The same principal geometry is used in the initial design with a flat detector and in the re-optimized design with a toroidal detector.}}
\label{optics}
\end{figure}

The general view of the telescope optical system is shown on Fig.\ref{optics}. The demonstrative example under consideration has the focal length of \textit{f’}=960 mm, the aperture of \textit{F/5.3}, and a rectangular field of view $\omega_x$x$\omega_y$=4$^{\circ}$x4$^{\circ}$.

All the three mirrors are hyperbolic sections with a common axis and they are used off-axis in both aperture and field angle. The surface of each mirror is described by the conic section equation:

\begin{equation}
z(\rho)=\frac{\frac{1}{R}\rho^2}{1+\sqrt{1-(1+k)]\frac{\rho^2}{R^2}}}.
\label{eq:refname1}
\end{equation}

In contrast with a spherical surface the curvature varies with the coordinates and it depends on the direction. The curvature radii along X and Y (i.e. in the tangential and sagittal directions) are:

\begin{equation}
R_y(x,y)=\frac{\big[1+(\frac{\partial z}{\partial y})^2 \big]^{3/2}}{\frac{\partial^2 z}{\partial y^2}};
R_x(x,y)=\frac{\big[1+(\frac{\partial z}{\partial x})^2 \big]^{3/2}}{\frac{\partial^2 z}{\partial x^2}}.
\label{eq:refname2}
\end{equation}

It is then possible to calculate the Petzval curvature radii along X and Y separately

\begin{equation}
R_{fy(x)}=\frac{1}{\sum_{i=1}^{k}\frac{2}{R_{y(x)}}}
\label{eq:refname3}
\end{equation}

By applying the equations (\ref{eq:refname2})-(\ref{eq:refname3}) to the current design, we obtain for the focal surface radii $R_{fy}=-1699.7 mm;R_{fx}=-1134.4 mm$.  
Then we re-optimize the entire design with the standard tools implemented in the Zemax software varying the radii, conic constants and distances, but keeping the focal length and the principal geometry to avoid obscuration. After the optimization we obtain $R_{fy}=-1808.2 mm;R_{fx}=-1188.9 mm$. \textbf{The mirrors surfaces remain standard second order aspheres, though their exact shape slightly changes:the root mean square (RMS) deviations from the best fit spheres (BFS) in the initial design  are 53.9, 25.6 and 60.9 $\mu m$ for the primary, secondary and tertiary, respectively; while in the re-optimized design they are 47.8, 22.1 and 60.9 $\mu m$.} Further we demonstrate the image quality achieved in this design.

\subsection{Image quality analysis}

\begin{figure}[h]
\centering
\fbox{\includegraphics[width=0.9\linewidth]{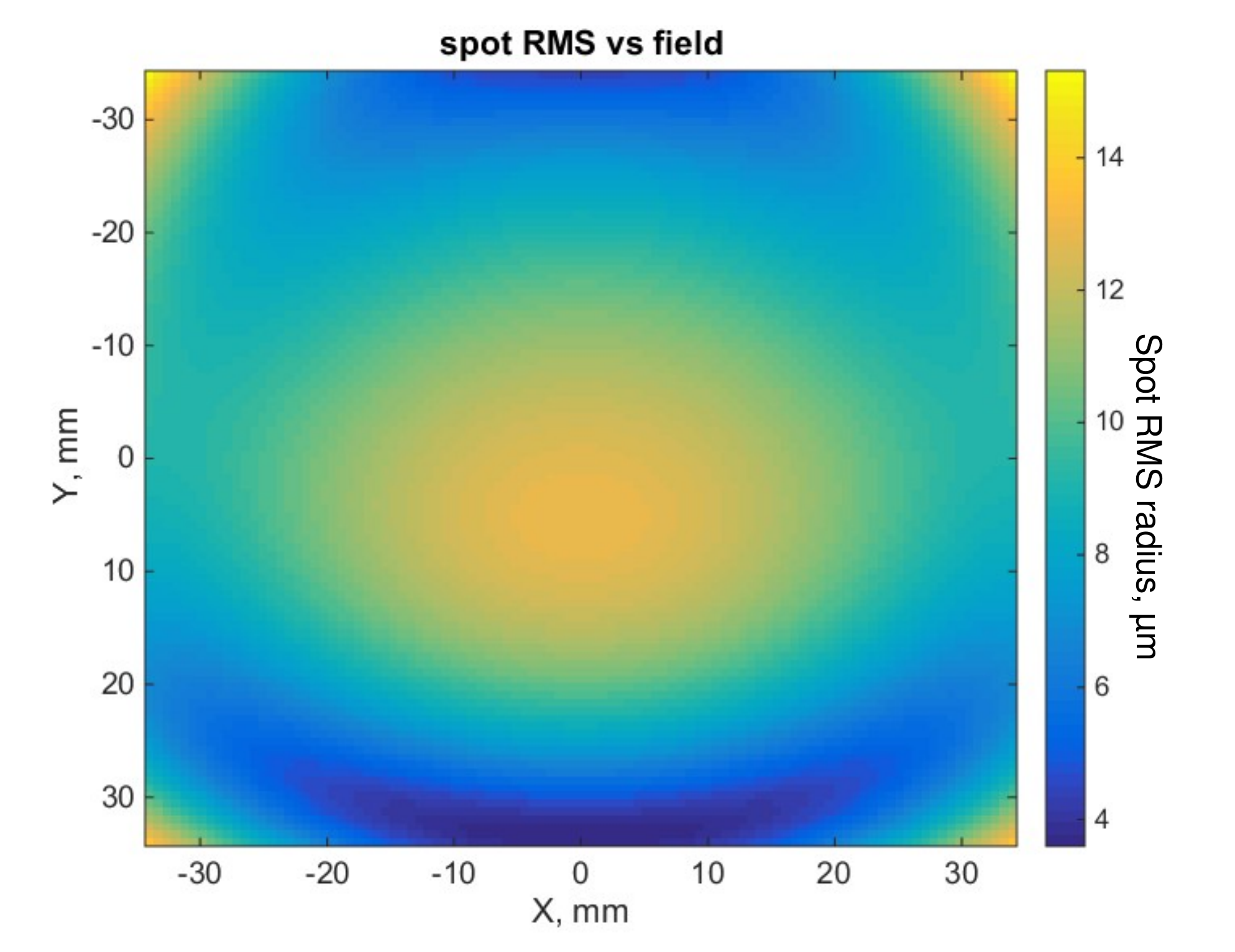}}
\caption{Field map of the telescope - the spot RMS radius vs. \textbf{linear coordinates on the detector surface.} The maximum value is 15.3 ${\mu}m$ , the mean is 8.9 ${\mu}m$.}
\label{map}
\end{figure}

\begin{figure}[h]
\centering
\fbox{\includegraphics[width=0.9\linewidth]{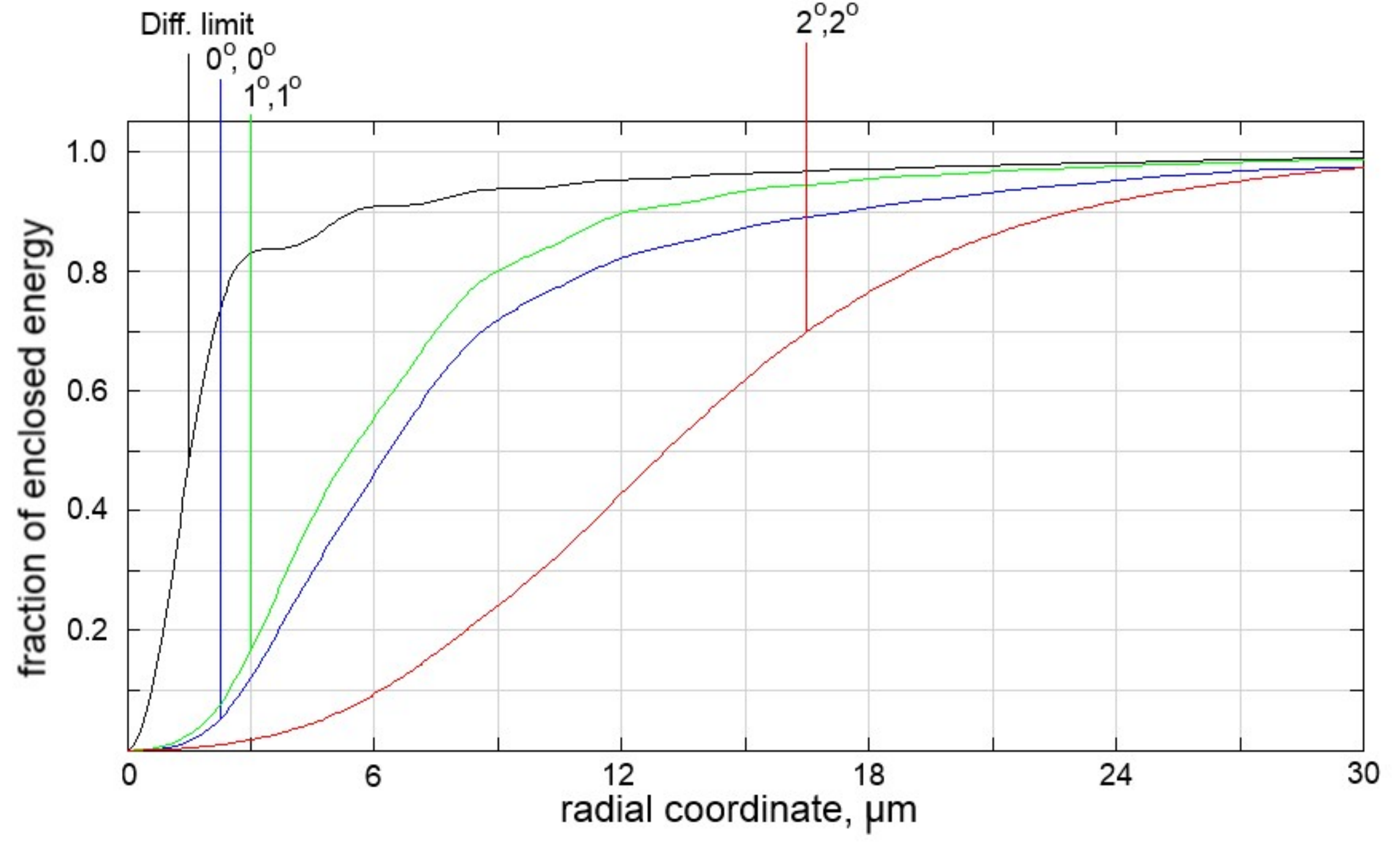}}
\caption{Energy concentration plot for the telescope with toroidal detector\textbf{ at 3 control field points. All the values approach the diffraction limit at 30 $\mu m$ }(the reference wavelength is 550 nm).}
\label{enc}
\end{figure}

To asses the image quality we consider the map of spot \textbf{RMS} radii depending on the field (see Fig.\ref{map}) and the encircled energy plot (see Fig.\ref{enc})

\begin{figure}[h]
\centering
\fbox{\includegraphics[width=0.9\linewidth]{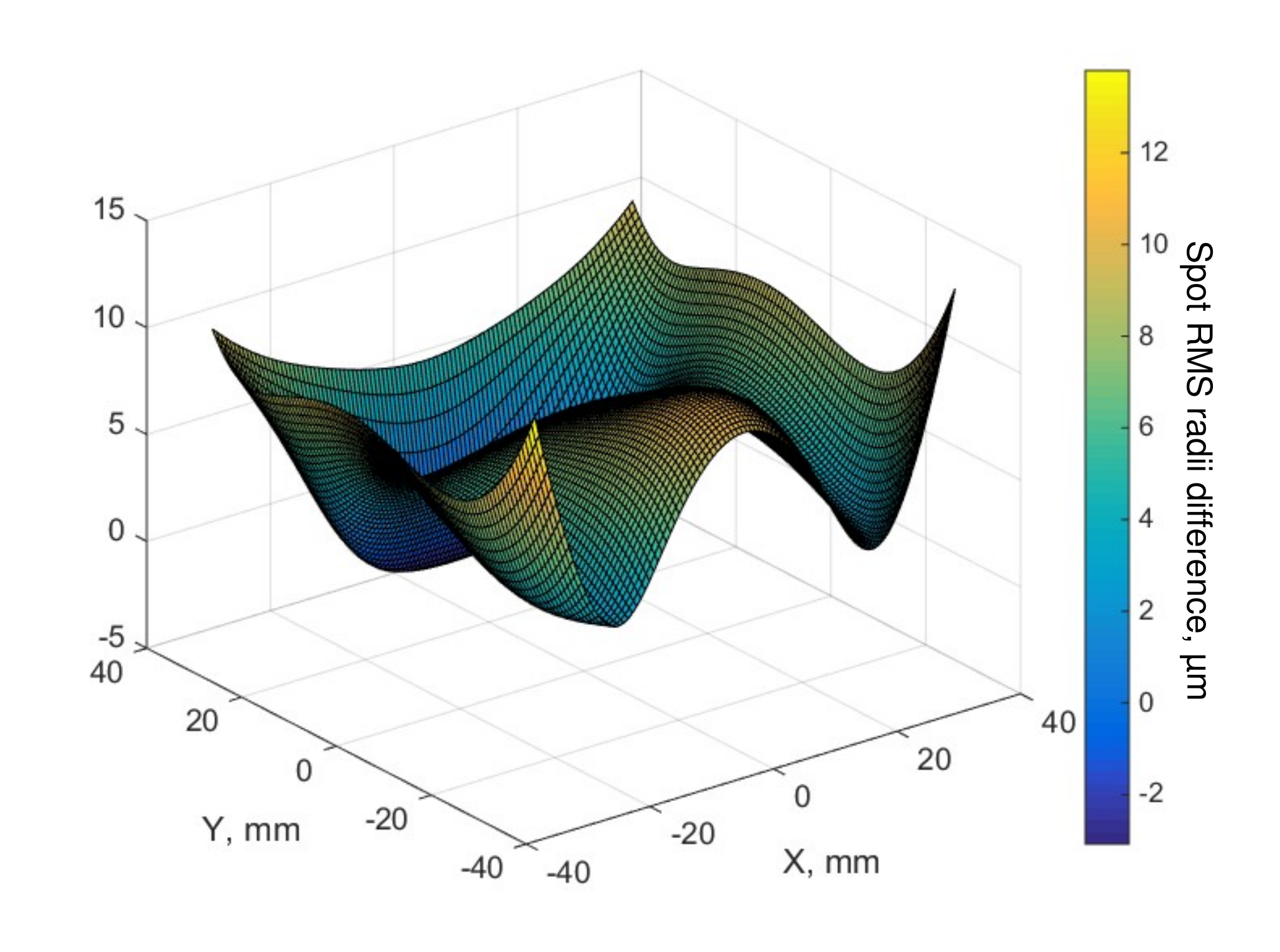}}
\caption{Difference of the spot RMS radii between the telescope designs with flat and toroidal detectors.\textbf{ Similarly, the map is given in the linear coordinates on the detector surface.} The maximum value is 13.8 ${\mu}m$, the mean is 5.4 ${\mu}m$. }
\label{compare}
\end{figure}

In order to demonstrate the gain obtained by use of the toroidal detector we consider the difference between the spot radii maps for the initial design with a flat detector and the new one (Fig.\ref{compare}). The advantage of the toroid-based design is substantial.

\section{Fabrication of the toroidal detector}

We assume that the curved detector is manufactured by warping a back-thinned detector chip attached to a deformable substrate \citep{Jahn16}. In this particular case we consider a variable thickness distribution (VTD) plate deformed by an uniform pressure and reaction at the edge \citep{Lemaitre76}, \citep{Lemaitre09}. This case leads to a cycloid-type VTD, which remains relatively simple when extending it to the case of toroidal shape. There are other options like deformation by a center force and an edge reaction or center force combined to a uniform load. Using a constant thickness plate controlled by a number of individual actuators is also possible. However, this option is out of preference due to complexity of the actuators calibration and possible generation of high-order deformations \citep{Laslandes12}. 

\subsection{Theoretical elasticity analysis}

For the chosen geometry type we can use the following equation to define the VTD plate profile:

\begin{equation}
t(r)=-{\Big[3(1 - \nu)\frac{qR_f}{Ea}\Big(1 - \frac{r^2}{a^2}\Big)}\Big]^{1/3}a
\label{eq:refname4}
\end{equation}

Here $\nu$ is the plate material Poisson ratio, \textit{E} is the Young's modulus, \textit{q} is the uniform load value and \textit{a} is the plate semi-diameter. 
The equation (\ref{eq:refname4}) is applied separately for the X and Y directions. Besides, the condition should be matched to keep the same center thickness: 

\begin{equation}
a_x=\sqrt{\frac{R_{fy}}{R_{fx}}}a_y
\label{eq:refname5}
\end{equation}

\begin{figure}[ht]
\centering
\fbox{\includegraphics[width=0.9\linewidth]{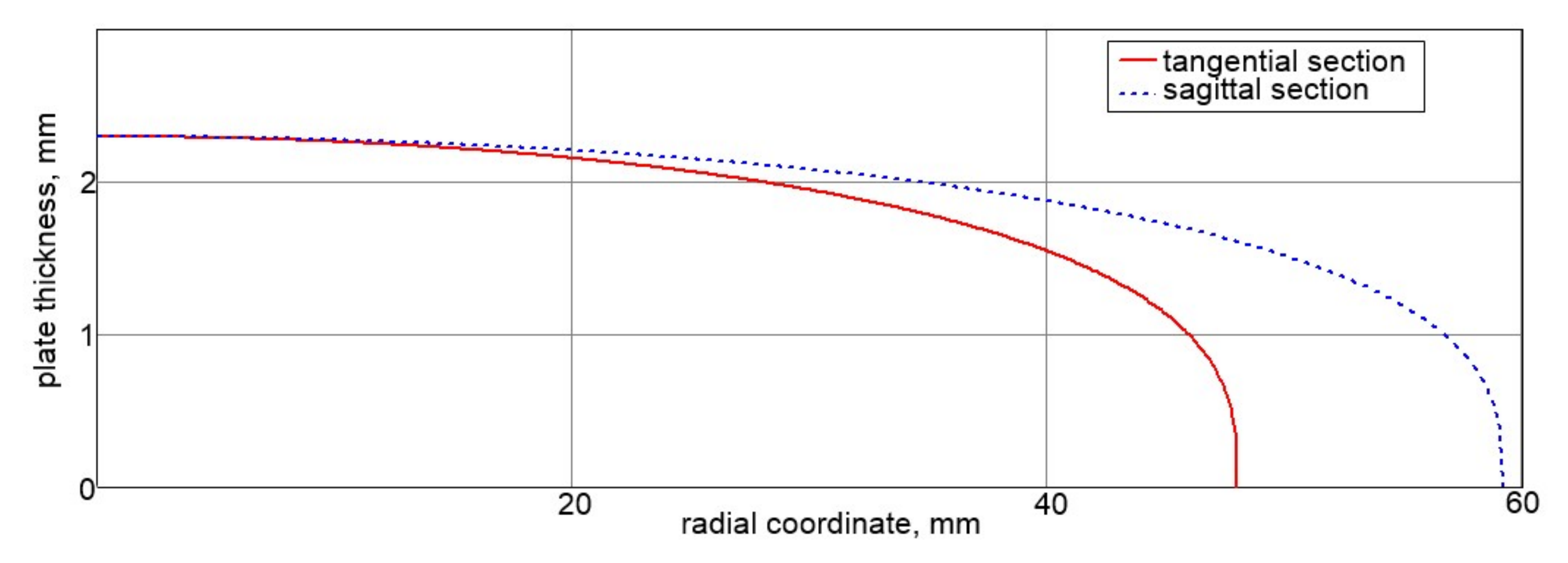}}
\caption{Theoretical thickness distribution for the deformable plate in the tangential and sagittal sections. \textbf{Further they are used to construct the variable thickness distribution plate as the generating curves.} }
\label{thickness}
\end{figure}

Assuming that the substrate is made out of an aluminum alloy (see the mechanical properties, for example, in \citep{Lingaiah03}) and that the uniform load is $10^5$ Pa, we obtain the profiles shown on Fig. \ref{thickness}.  

\subsection{Finite element modeling}

To finalize the elasticity design we built a solid 3D model of the bending \textbf{mechanism}, where the deformable substrate is connected to a bulky ring by a thin collar. The anamorphic substrate was obtained by combination of the two analytical profiles with an elliptical generating curve. \textbf{More details about the approach to design an analogous bending mechanism can be found in \citep{Ferrari98}.} We must note that this shape can be manufactured by Computer Numerical Control (CNC) milling or by 3D printing \citep{Hugot16}. 
\textbf{The silicon chip is thinned down to 100 ${\mu}m$ and attached on top of the substrate with a silicone glue. }       

\begin{figure}[ht]
\centering
\fbox{\includegraphics[width=0.9\linewidth]{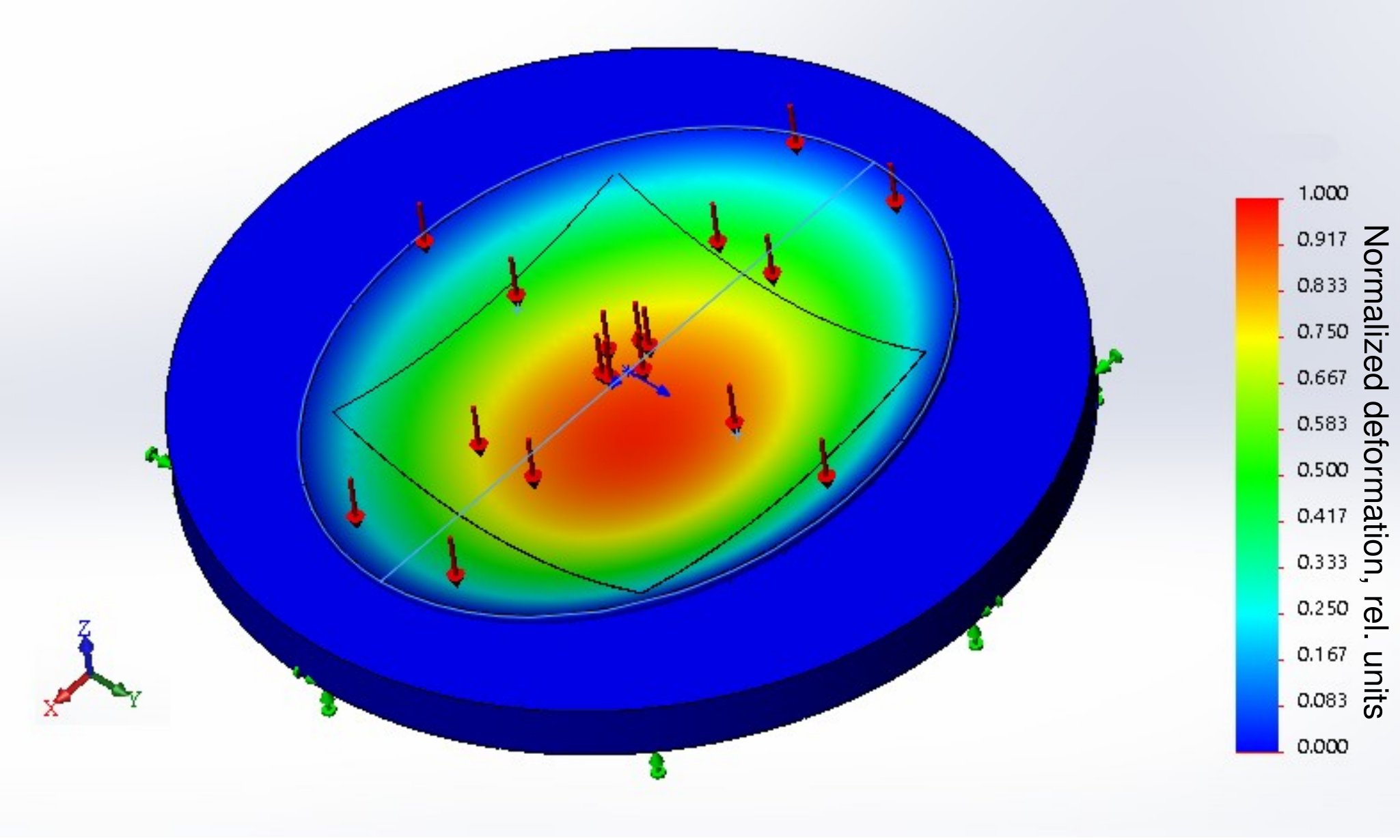}}
\caption{\textbf{Results of for finite element analysis. The color map represents normalized deformation of the plate, the green arrows indicate fixed geometry and the red ones are for the bending load.}}
\label{FEA}
\end{figure}

The detector bending was modelled with a FEA software. In order to account for the additional layers influence and minor changes in the plate geometry the pressure was changed. The final adjusted value is $1.12 10^5$ Pa. The results of the FEA simulation are shown on Fig. \ref{FEA}. 

\begin{figure}[ht]
\centering
\fbox{\includegraphics[width=0.9\linewidth]{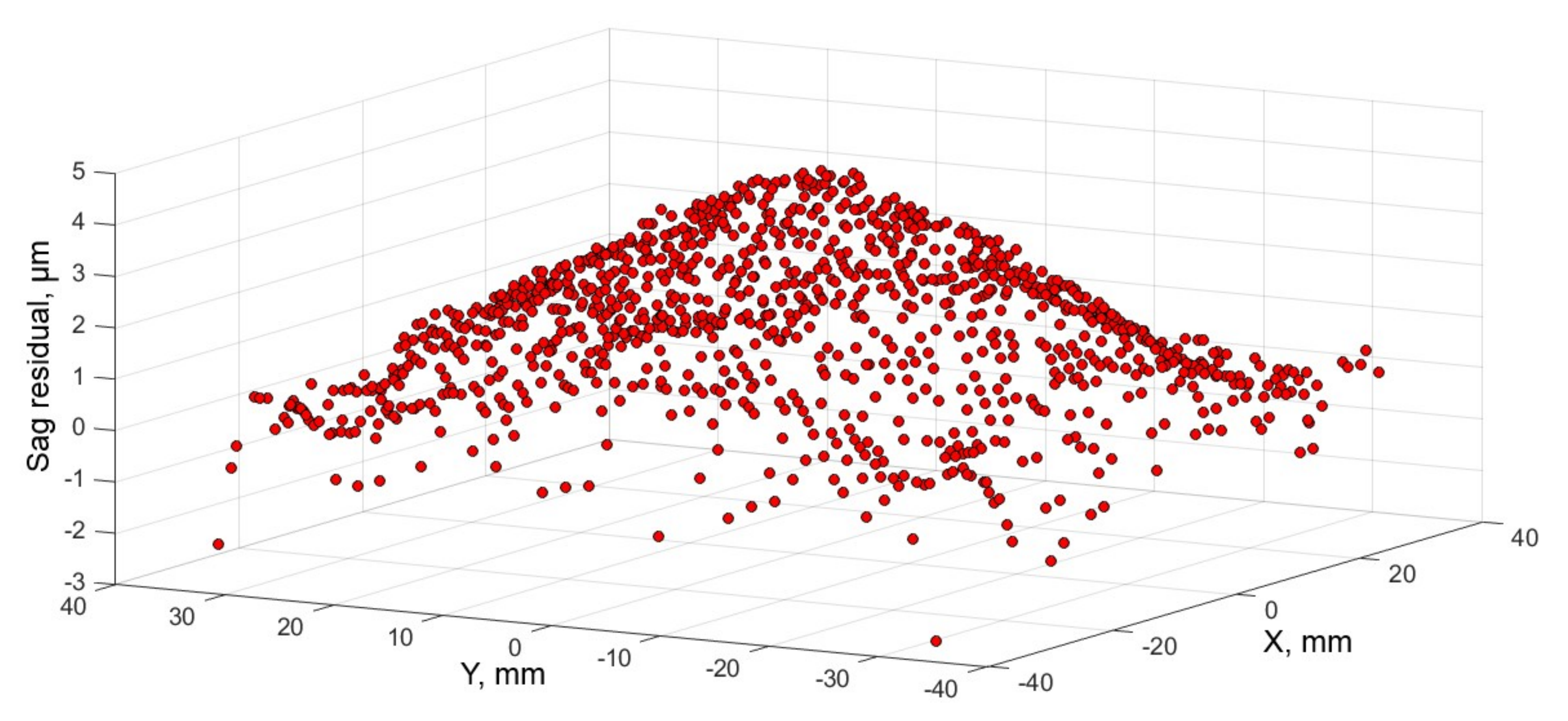}}
\caption{Residual sag error of the curved \textbf{toroidal} detector shape: the difference between target toroidal shape and the FEA results (PTV=7.4 ${\mu}m$, RMS=1.9 ${\mu}m$). \textbf{This deviation cannot affect the image quality.}}
\label{residuals}
\end{figure}

When comparing the obtained shape with the required toroid, we find that the residual discrepancy is quite small with an RMS value lower than 2.0 ${\mu}m$ \textbf{(see the deviation plot in Fig.\ref{residuals}).} This difference cannot affect the image quality.

\section{Conclusions}

We proposed an approach to design optical systems with toroidal detectors. It was successfully shown on the example of an off-axis TMA telescope that re-optimization with a toroidal detector makes it possible to achieve a notable gain in the image quality (up to 40 \% in terms of the spot RMS radius). The optical scheme has the same principal geometry and uses the similar aspherical mirrors as the initial one. At the same time, the toroidal detector can be manufactured with use of a bending \textbf{mechanism} designed on the basis of analytical computations. 
\textbf{In conclusion we would like to mention a few of the key potential applications of this design approach.} It can be applied for existing off-axial optical designs like the one presented here. Also it can be extremely useful when creating new, unobscured, decentered imaging systems. Other classes of optical schemes without rotational symmetry like spectrographs and systems with image slicers \citep{Henault03} can benefit from using toroidal detectors. However, one should perform a thorough aberration analysis before performing the re-optimization. If the image quality is dominated by other aberrations (especially coma-type aberration) the advantage of use of the toroidal detector can be negligible.

The authors would like to acknowledge the European comission for funding this work through the Program H2020-ERC-STG-2015 – 678777 of the European Research Council, as well as the French Research Agency through the program LabEx FOCUS ANR-11-LABX-0013.
The authors thank Bertrand Chambion, David Henry and Christophe Gaschet from CEA-LETI (Grenoble) for their crucial contribution to the curved detectors development.

\bigskip

\renewcommand{\bibname}{References}

\bibliographyfullrefs{sample}

\renewcommand{\bibname}{}

\ifthenelse{\equal{\journalref}{aop}}{%
\section*{Author Biographies}
\begingroup
\setlength\intextsep{0pt}
\begin{minipage}[t][6.3cm][t]{1.0\textwidth} 
  \begin{wrapfigure}{L}{0.25\textwidth}
    \includegraphics[width=0.25\textwidth]{john_smith.eps}
  \end{wrapfigure}
  \noindent
  {\bfseries John Smith} received his BSc (Mathematics) in 2000 from The University of Maryland. His research interests include lasers and optics.
\end{minipage}
\begin{minipage}{1.0\textwidth}
  \begin{wrapfigure}{L}{0.25\textwidth}
    \includegraphics[width=0.25\textwidth]{alice_smith.eps}
  \end{wrapfigure}
  \noindent
  {\bfseries Alice Smith} also received her BSc (Mathematics) in 2000 from The University of Maryland. Her research interests also include lasers and optics.
\end{minipage}
\endgroup
}{}

\end{document}